\documentstyle[prl,aps,epsfig,amssymb,citesort]{revtex}
\def\B.#1{{\bbox{#1}}}

\def\r{{\mbox{\boldmath$r$}}}

\def\unitr{{\mbox{\boldmath$\hat r$}}}

\def\unitz{{\mbox{\boldmath$\hat z$}}}

\def\v{{\mbox{\boldmath$v$}}}

\def\unitr{{\mbox{\boldmath$\hat r$}}}
\def\unitz{{\mbox{\boldmath$\hat z$}}}

\def\la{{\langle}}
\def\ra{{\rangle}}

\renewcommand{\S}{{\cal S}}
\renewcommand{\B}{{\cal B}}

\newcommand{\rb}{Rayleigh-B\'enard }
\newcommand{\lp}{\left(}
\newcommand{\rp}{\right)}
\newcommand{\be}{\begin{equation}}
\newcommand{\ee}{\end{equation}}
\newcommand{\bea}{\begin{eqnarray}}
\newcommand{\eea}{\end{eqnarray}}

\begin{document}
\twocolumn[\hsize\textwidth\columnwidth\hsize\csname
@twocolumnfalse\endcsname
\title{Universality of anisotropic fluctuations from numerical simulations of turbulent flows }
\author{L. Biferale$^{1,4}$, E. Calzavarini$^{2}$, F. Toschi$^{3,4}$ and R. Tripiccione$^{2}$}
\address{$^1$ Dipartimento di Fisica, Universit\`a ``Tor Vergata'', Via della Ricerca Scientifica 1, I-00133 Roma, Italy}
\address{$^2$ Universit\`a degli Studi di Ferrara, and INFN, Sezione di Ferrara, Via Paradiso 12, I-43100 Ferrara, Italy }
\address{$^3$ CNR, Istituto per le Applicazioni del Calcolo, Viale del Policlinico 137, I-00161, Roma, Italy}
\address{$^4$ INFM, Unit\`a di Tor Vergata, Via della Ricerca Scientifica 1, I-00133 Roma, Italy}
\maketitle
\begin{abstract}
We present new results from a direct numerical simulation of a three
dimensional {\it homogeneous} \rb system (HRB), i.e.  a convective
cell with an imposed linear mean temperature profile along the
vertical direction.  We measure the SO(3)-decomposition of both
velocity structure functions and buoyancy terms. We give a dimensional
prediction for the values of the anisotropic scaling exponents in this
\rb systems.  Measured scaling does not follow dimensional estimate,
while a better agreement can be found with the anisotropic scaling of
a different system, the {\it random}-Kolmogorov-flow (RKF) \cite{rkf2}.
Our findings support the conclusion that scaling properties of
anisotropic fluctuations are {\it universal}, i.e. independent of the
forcing mechanism sustaining the turbulent flow.
\vskip0.3cm

\end{abstract}
] Small scales turbulent statistics is a challenging
open problem for both theoretical and experimental studies in
hydrodynamical systems \cite{frisch}. Typical questions are connected
to the understanding of the {\it universality} issue, i.e. to which
extent small-scale turbulent fluctuations are statistically
independent of the large-scale set-up used to inject energy in the
flow. Robustness of small-scale physics cannot be exact.  For
instance, different forcings may inject large-scale different
anisotropic fluctuations, which must have some direct/indirect
influence on small-scale statistics.

A first strong requirement for
{\it universality} to hold is therefore that large-scale anisotropic
fluctuations becomes more and more sub-leading by going to smaller and
smaller scales. In other words, at scales small enough, the {\it
omnipresent and universal} isotropic fluctuations must be the leading
statistical components. Such a requirement is always observed in both
experiments and numerical simulations, although some subtle effects
may show up due to the existence of anomalous anisotropic scaling (see
\cite{rkf1,rkf2,bv00,sw00} for a detailed discussion of this
issue). Another important question which must be asked about {\it
universality} of small scales statistics, is connected to the
anisotropic components on their own, independently on their comparison
with the isotropic ones. In particular, it is  important to
understand whether the anisotropic components of any turbulent
correlation functions have a scaling behavior characterized by {\it
universal} exponents or not, in the limit of high Reynolds numbers. \\

In this Letter we present first results of an attempt to study the
small-scale anisotropic behavior of a {\it homogeneous} three
dimensional \rb system (HRB), i.e.  a convective cell with fixed
linear mean temperature profile along the vertical direction. The main
focus of our analysis is a comparison between the statistical behavior
of  HRB system with a completely different anisotropic flow, a
random-Kolmogorov-flow (RKF) \cite{rkf1,rkf2}. From the comparison, we
show that the two systems have almost indistinguishable, in the limit
of our numerical resolution, small-scale anisotropic (and isotropic)
scalings, i.e. we find a high degree of small-scale universality for
all measurable anisotropic components. This result is particularly
relevant because its validity is only possible if HRB has {\it
anomalous} (to be defined below in details) anisotropic small-scale
fluctuations.

This Letter is organized as follows. First we briefly discuss the
physics of  HRB flow and the details of our numerical
simulations. Second, we review the technique of SO(3) decomposition to
disentangle different anisotropic contributions to any velocity
correlation functions. We then present our numerical results on the
HRB.

We first show that the observed anisotropic scaling is {\it
anomalous}, i.e. it does not follow the dimensional predictions than
can be derived by an analysis of the equation of motion.  We then
address in details small-scale {\it universality} by making the
comparison between  HRB and RKF anisotropic properties, the
central point of the present Letter.
\begin{figure}[!bh]
\hskip -.4cm\epsfig{file=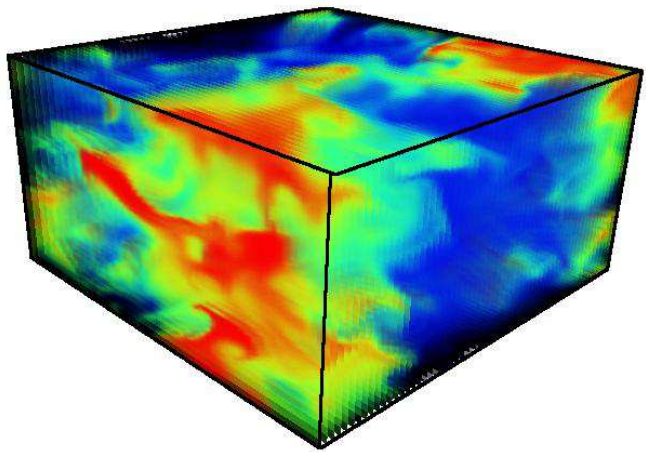,width=\hsize}
\caption{\label{fig0} A snapshot of the temperature
 fluctuations in the 3D HRB system. The system is fully periodic.
 Notice the presence of typical convective structures {\protect \cite{kada}}
 as the neat
 plume on the bottom/left of the picture. }
\end{figure}
\noindent
 
An  Homogeneous \rb system (HRB) is a convective cell with fixed
linear mean temperature profile along the vertical direction.  The
flow is obtained by decomposing the temperature field as the sum of a
linear profile plus a fluctuating part, $T(x,y,z;t)= T'(x,y,z;t) + (\Delta
T/2 - z \Delta T/H) $, where $H$ is the cell height and $\Delta T$
the background temperature difference.  The evolution of a HRB system can
be described by a modified version \cite{lohse2} of the Boussinesq system
\cite{kada}:
\bea
\label{omeq1} \partial_t \v &+& \lp\v\cdot\nabla\rp \v =
 -\nabla p + \nu \nabla^2 \v + \alpha g T' {\unitz}\\
\label{omeq2} \partial_t T' &+& \lp\v\cdot\nabla\rp T' =
  \kappa \nabla^2 T' - {{\Delta T} \over H} v_z.
\eea
Fully periodic boundary conditions are used for the velocity field,
$\v$, and temperature, $T'$, fields.\\ For large Rayleigh numbers, HRB
shows a turbulent convective dynamics with absence of both viscous and
thermal boundary layers \cite{lohse2}.  The Bolgiano scale, $L_B \equiv
\epsilon^{5/4} N^{-3/4} (\alpha g)^{-3/2}$, is of the order of the
integral scale of the cell (H), hence temperature fluctuations have a
leading  role only at the largest scales in the system.  This has been
already shown in a similar simulation \cite{orsz}, and is consistent
with the picture presented in \cite{rb}.  

The main advantage of the HRB system is that the intrinsic homogeneity
along the three directions allow a systematic study of scaling
properties without spurious (non-homogeneous) effects, always present
in standard
\rb systems with boundary layers.\\

In order to asses the importance and properties of anisotropic
components for any correlation function it is necessary to make a
decomposition on the complete basis of the SO(3) group
\cite{alp99}. In particular, in the following, we  are mainly
interested in the SO(3) decomposition of scalar quantities, as for the
case of velocity longitudinal structure functions, $ S^{(p)}(\r) =
\la\left[\lp\v\lp \r \rp-\v\lp 0 \rp\rp\cdot
\unitr\right]^p\ra\,$:
\begin{equation}
S^{(p)}(\r) = \sum_{j=0}^{\infty}\sum_{m=-j}^{j} \S^{(p)}_{jm}(r)
Y_{jm}(\unitr );
\label{so3_sf}
\end{equation}
where the indices $(j,m)$ label the total angular momentum and its
projection on a reference axis of the spherical harmonics
$Y_{jm}(\unitr)$, respectively (see \cite{alp99,rkf1} for more
details). The physics is hidden in the projections,
$\S^{(p)}_{jm}(r)$. We are interested to measure what are the scaling
properties (if any) of each projections on different anisotropic
sectors: \be
\S^{(p)}_{jm}(r) \sim c_{jm} r^{\xi^j(p)}.
\ee
where we have assumed, on the basis of theoretical \cite{alp99} and
numerical \cite{rkf1} evidences, that the scaling exponents do not
depend on the $m$ index.  To go back to the {\it universality} issue
discussed at the beginning; we expect that the coefficients $c_{jm}$
are strongly dependent on the anisotropic properties of the
large-scale physics while the values of the scaling exponents,
$\xi^j(p)$, must enjoy a much higher degree of {\it universality}.  In
other words, whether a particular sector is alive, $|c_{jm}|>0$, or
not, $c_{jm}=0$, depends on the forcing; while, once that sector is
switched-on, the way it propagates to small-scale is
forcing-independent. This picture can be proved on a rigorous basis
for some problems of scalar/vector advection by Gaussian,
white-in-time, velocity fields (Kraichnan models
\cite{rev_sca}).\\
Concerning the SO(3) analysis let us notice that velocity structure
functions have even parity with respect to $\r$, therefore projections
on odd $j$ values vanish.  In the following, we consider also mixed
velocity and temperature structure functions which have, on the other
hand, dominant odd parity.\\ From equation (\ref{omeq1}) one may
easily write down the stationary equation for the second order
velocity structure functions; the extension of K\'arm\'an-Howarth
equation in the presence of a buoyancy term \cite{ko_rb}. The result
is, neglecting for simplicity tensorial symbols:
\bea
\label{general}
\la \delta v(\r)^3 \ra &\sim& \ \ \epsilon \ r \ + \ \alpha g\unitz r\
\cdot \ \la \delta T(\r)\ \delta v(\r)\ra\\ _{j = 0,1,\dots} &\ & \ \
_{j=0} \quad \ \ _{j=1 \quad \otimes \quad j=1,2,\dots} \nonumber \eea
where with $\epsilon$ we denote the energy dissipation and with, $\la
\delta v(\r)^3 \ra$ and $\la \delta T(\r)\ \delta v(\r)\ra$, the
general third-order velocity correlation and temperature-velocity
correlation, respectively. The two terms on the r.h.s. of equation
(\ref{general}) are called respectively the energy-dissipation term
and buoyancy term.  In (\ref{general}) we report for each term the
value of its total angular momentum, $j$.  Let us notice that the
energy dissipation term in (\ref{general}) has a non-vanishing limit,
for high Reynolds numbers, only in the isotropic sector, $j=0$. On the
other hand, the buoyancy coupling, $\alpha g\unitz$, brings only
angular momentum $j=1$. Due to the usual rule of composition of angular
momenta we have that the buoyancy term, $\alpha g \unitz \cdot \la
\delta T(\r)\ \delta v(\r)\ra$, has a {\it total} angular momentum
given by the rule: $j_{tot}= 1 \otimes j = \{j-1,j,j+1\}$.  Using the
angular momenta summation rule for $j$ we can decompose the previous
equation obtaining the following dimensional matching, in the
isotropic sector: \be
\label{iso}\la \delta v(r)^3 \ra_{j=0} \sim  \epsilon\ r +  
\alpha g\ {\mathbf\hat{z}} r\ \la \delta v(r) \ \delta
T(r)\ra_{j=1} + \dots 
\label{aniso1}
\ee
and in the anisotropic sectors, $j >0$:
\be
\la \delta v(r)^3 \ra_{j} \sim  \ \ \alpha g\ {\mathbf\hat{z}} r\ \la \delta v(r) \ \delta T(r) \ra_{(j-1)} + \dots 
\label{aniso}
\ee
where only dominant contributions are reported.

In the isotropic sector the buoyancy term is sub-dominant with respect
to the dissipation term at scales smaller than the Bolgiano length, $
r< L_B$ \cite{note1}. Therefore, in our simulation the isotropic
velocity fluctuations are closer to the typical Kolmogorov scaling,
$\delta v(r) \sim r^{1/3}$, rather than to the Bolgiano-Obhukhov
scaling, $\delta v(r) \sim r^{3/5}$.\\ Let us now focus on the
anisotropic sectors.  Equation (\ref{aniso}) is the simplest {\it
dimensional prediction} one can derive for this system consistently
with the anisotropic properties of the buoyancy term, sector by
sector.  It plays a key role in the following because we will show
that the observed anisotropic scaling in our HRB system differ from
the matching (\ref{aniso}), i.e. we measure anomalous anisotropic
scaling exponents.

Our HRB simulation was performed using a Lattice Boltzmann scheme,
with spatial resolution of $240^3$. We stored roughly 270 statistical
independent configurations.  The Prandtl number for the simulation is
equal to unit, and the Rayleigh number $Ra =(\alpha g \Delta T H^3
)/(\nu \kappa)=1.38 \cdot 10^7$.  Measured Bolgiano scale is $L_B \sim
370$, roughly one and half the cell size, while $\alpha g$ used in the
equation of motion (\ref{omeq1}) is $2\cdot 10^{-3}$.  A typical
snapshot of the temperature field is shown in Figure
\ref{fig0}. Notice the well detectable structures typical of all other
\rb cell \cite{kada,lohse1,siggia,sergio}.  In particular, there is a
beautiful hot plume on the central bottom/left part of the picture.

We now present our numerical results.  In order to check the
small-scale properties of the HRB system we have carried out the
SO(3) decomposition of both longitudinal velocity structure functions
(\ref{so3_sf}) up to order $p=6$ and of the simplest set of
buoyancy-like terms made of $q$ velocity longitudinal increments and
of one temperature increment, $B^{(q,1)}(\r) =
\la\left[\lp\v\lp\r\rp-\v\lp 0 \rp\rp\cdot \unitr\right]^{q}\lp T\lp
\r \rp-T\lp 0 \rp\rp \ra\,$:
\begin{equation}
B^{(q,1)}(\r) = \sum_{j=0}^{\infty}\sum_{m=-j}^{j}
\B^{(q,1)}_{jm}(r) Y_{jm}(\unitr ).
\label{so3_b}
\end{equation}
The dimensional matching made in (\ref{aniso}) can be extended to any
order of correlation function, giving, in terms of the velocity and
buoyancy projections, the leading scaling contribution:
\be
  \S^{(p)}_{jm}(r) \sim r \B^{(p-2,1)}_{j-1,m}(r).
\label{comp_dim}
\ee
Denoting with $\chi^j(q,1)$ the anisotropic scaling properties of
the buoyancy terms, $\B^{(q,1)}_{j-1,m}(r) \sim r^{\chi^j(q,1)}$ we
get the dimensional estimate linking velocity and buoyancy scaling:
\be
\xi_d^j(p) = 1 + \chi_d^{j-1}(p-2,1)
\label{dim_exp}
\ee
where we have added a subscript $d$ to remind the reader that it is
the result of a dimensional matching.

\begin{figure}[!h]
\epsfig{file=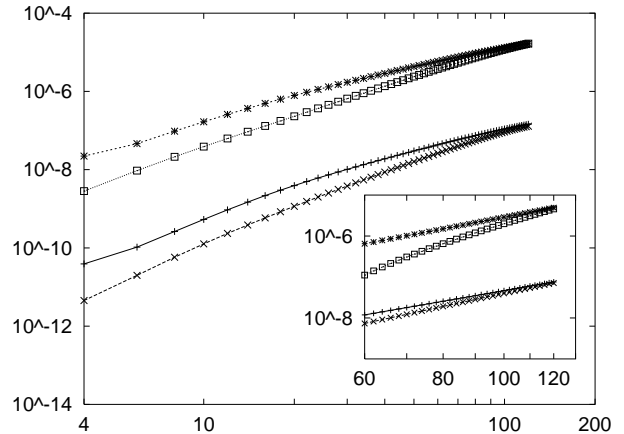,width=\hsize}
\caption{ \label{fig1} Log-log plot of quantities entering in the
dimensional matching (\ref{comp_dim}) for some anisotropic sectors
{\it vs} $r$. Two top curves refer to $\S^{(p)}_{j,0}(r)$ ($\ast$) and
to the buoyancy term $r\B^{(p-2,1)}_{j-1,0}(r)$ ($\square$)
with $p=3$ and $j=4$. The two bottom curves refer to the case $p=5$
with symbols, ($+$) for the velocity projection and ($\times$) for the
buoyancy term.  Inset: same for $j=6$.  Notice that the buoyancy term
decays faster in all cases.  In both plots, buoyancy terms are shifted along $y$-axis for the sake of presentation and we limited the $r$-range
extension to those values with a statistically significant signal.}
\end{figure}

Let us first discuss the issue of anisotropic {\it anomalous} scaling
by making a comparison between the numerical measurements and the
dimensional estimate (\ref{dim_exp}).  
In Fig. \ref{fig1} we show a comparison between
the velocity and buoyancy $j=4$ projections for $p=3,5$. 

 The inertial scaling measured for the projection of the velocity
structure function, $\S^{(p)}_{jm}(r)$, is more intermittent than the
corresponding buoyancy term, $r\B^{(p-2,1)}_{j-1,m}(r)$. In other
words, the observed velocity scaling is different from the dimensional
matching derived from the equations of motion: it is {\it anomalous}.
This result holds for all measurable orders also for $j=2$ and for
$j=6$ sector ($j=6$ is shown in the inset).  

\begin{figure}[!t]
\hskip -.4cm\epsfig{file=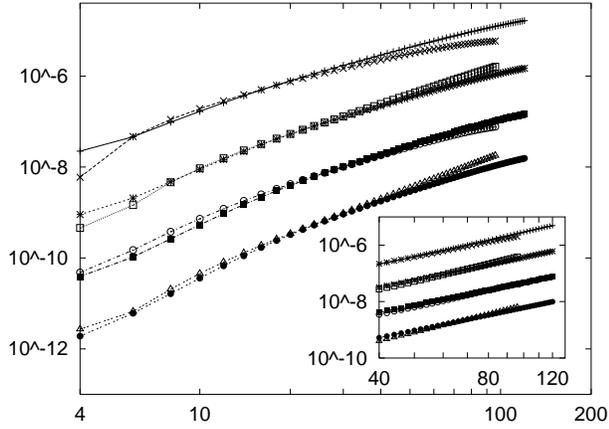,width=\hsize}
\caption{\label{fig:2}  Log-log plot of anisotropic $j=4$ projections $\S^{(p)}_{j,0}(r)$ 
{\it vs} $r$ for both HRB and RKF flows.  Each couple of --almost--
coinciding curves refer to the comparison of HRB and RKF projections
at a given order $p=3,4,5,6$ (top to bottom).  Inset: same for $j=6$.}
\end{figure}

Let us now discuss the other important issue of {\it universality}. We
argue that not only HRB has anomalous anisotropic scaling but also
that the measured behavior is indistinguishable from what measured in
the random-Kolmogorov-flow \cite{rkf1}.  The point is far from being
trivial and must not be underestimated. The HRB has an anisotropic
forcing, given by the buoyancy term, which acts at all scales, $\sim g
\unitz \delta T(r)$, i.e. there is also a direct inject of
anisotropies at small-scale, at variance with the RKF where the
forcing was only at large scales.  In Fig. (\ref{fig:2}) we show an
overall comparison of $\S^{(p)}_{jm}(r)$ measured on the HRB and on
the RKF \cite{rkf1}.  Comparison is limited to $j=4$ and $j=6$,
because RKF data from the simulation of \cite{rkf1} have a sign
inversion in the $j=2$ sector which makes comparison inconclusive.
As can be seen the agreement is quite satisfactory, except for the
very small scales, smaller then the viscous scale, where as usual the
SO(3) decomposition suffers from interpolation errors. The small
discrepancies at large scales are also to be expected: the inertial
properties of the two flows have to match quite different conditions
at large scale. 

The same comparison for $j=6$, shown in the 
inset, qualitatively supports the same result.\\

The fact that
inertial-scales fluctuations of both flows are almost
indistinguishable is the first important confirmation of the
universality of anisotropic fluctuations in sectors with
$j=4,6$. Similar conclusions can be drawn for the $j=2$ sector in
different experimental set-up
\cite{sw00,ks00,ref:98ADKLPS} 
(the only sector measurable, indirectly, in experiments).  

Let us conclude by summarizing the two main results of this
Letter. First, anisotropic fluctuations in \rb systems are {\it
anomalous}. Second, notwithstanding the direct influence of the
forcing mechanism at {\em all} small-scale, anisotropic fluctuations
are {\it universal}, i.e.  the small-scale dynamics is dominated by
{\it anomalous} fluctuations, coming from the self-organization of the
inertial evolution. Similar behavior is at the origin of anomalous
scaling in Kraichnan models of passive/vector advection as already
discussed in the introduction. In the latter case, one connects
rigorously the anomalous inertial scaling with the existence of {\it
zero-modes} of the inertial operator (see for example
\cite{ref:00ALPP,ref:00ABP} for a detailed analysis of anisotropic
scaling in passive advection of scalar and vector quantities,
respectively). Here, for Navier-Stokes equation, one may only stress
the striking similarities, without being able to push it to some
rigorous statement.  Concerning universalities for the isotropic
scaling of this \rb system, we notice that due to the large value of
the Bolgiano scale --of the order of the system size-- we expect to
observe small-scale isotropic fluctuations close to the usual
Kolmogorov 1941 scaling (plus intermittency).  This is indeed the
case. The Bolgiano-Obhukhov isotropic regime with $\delta v(r) \sim
r^{3/5}$ cannot be accessed from this simulation. In the framework of
the dimensional matching (\ref{aniso1}) the existence of a
Bolgiano-Obhukhov scaling in the isotropic sector corresponds to a
leading influence of the buoyancy term in the scaling range
\cite{sreeni}.

We acknowledge useful discussion with A. Lanotte.  This research
was supported in part by the EU under the Grant No. HPRN-CT 2000-00162
``Non Ideal Turbulence'' and by the INFN (through access to the
APEmille computer resources).  E. C. has been supported by Neuricam
spa (Trento, Italy) in the framework of a doctoral grant program with
the University of Ferrara.


\begin{thebibliography}{99}
\bibitem{note1}  The scale $r_B$ where $\epsilon
\simeq \alpha g\ \la \delta v(r_B) \ \delta T(r_B)\ra_{j=1}$ can be taken as 
 an alternative definition of the Bolgiano length, see \cite{rb}.
\end{thebibliography}
\end{document}